# Non-metric Generalizations of Relativistic Gravitational Theory and Observational Data Interpretation

A. N. Alexandrov, I. B. Vavilova, V. I. Zhdanov

*Astronomical Observatory,Taras Shevchenko Kyiv National University*
*3 Observatorna St., Kyiv 04053 Ukraine*

**Abstract**

We discuss theoretical formalisms concerning with experimental verification of General Relativity (GR). Non-metric generalizations of GR are considered and a system of postulates is formulated for metric-affine and Finsler gravitational theories. We consider local observer reference frames to be a proper tool for comparing predictions of alternative theories with each other and with the observational data. Integral formula for geodesic deviation due to the deformation of connection is obtained. This formula can be applied for calculations of such effects as the bending of light and time-delay in presence of non-metrical effects.

## 1. Introduction

General Relativity (GR) experimental verification is a principal problem of gravitational physics. Research of the GR application limits is of a great interest for the fundamental physics and cosmology, whose alternative lines of development are connected with the corresponding changes of space-time model.

At present, various aspects of GR have been tested with rather a high accuracy (Will, 2001). General Relativity is well tested experimentally for weak gravity fields, and it is adopted as a basis for interpretation of the most precise measurements in astronomy and geodynamics (Soffel et al., 2003). Considerable efforts have to be applied to verify GR conclusions in strong fields (Will, 2001).

For GR experimental verification we should go beyond its application limits, consider it in a wider context, and compare it with alternative theories. The basis of such a "theory of theories" was founded in 1960-th, when general requests to the gravity theory have been analyzed and a common approach to metric theories has been elaborated (Will, 1981, 2001). And the thought that the only metric theories, including GR, correspond adequately to the reality has been dominated. Nevertheless, development of gravitational theory, first of all, attempts of unification with the gauge approach and quantum theory maintain a steady interest to non-metric generalizations (see e.g. Haugan, Lammerzahl, 2001).

On the other hand, the deviations from metricity, if they exist, must be very small and at present we can say only about the upper limitations on values of corresponding parameters. However, these deviations may become large in strong gravitational fields and/or on cosmological scales.

For quantitative comparison of various gravity theories we must have a common scheme for the experimental data interpretation. In metric theories the Einstein equivalence principle provides such a common basis. When the non-metric generalizations have to be considered, there is a necessity to develop the generalized theory of reference frames and observables. It is important to remind that a proper time, angle between directions, local relative velocities and accelerations are the main space-time observables. Also, important information comes from photometrical, spectral and polarimetric measurements allowing us to determine distances and velocities of distant objects.

In this work we analyze a possibility to weaken the metric postulates and outline a field of GR generalizations, which fulfill these weakened postulates. We propose to use theory of local observer reference frames as a common basis for interpretation of the observational data and demonstrate its efficacy (Pyragas et al., 1995). At the end we consider a differential-geometrical problem, which is useful for subsequent applications: this concerns with perturbations of geodesics of Riemannian space induced by affine and Finsler deformations of connection.

## 2. Reliability criteria of gravity theory and metricity postulates

We remind the basic principles applied for the analysis of GR generalizations (Will, 1981). Any viable theory of gravity must satisfy *criteria of completeness, self-consistency, and correspondence with special relativity and Newton theory*. The first two criteria are of a general nature relevant to the theory of knowledge. Their fulfillment in the class of metric theories is closely connected with the universality of gravity interaction, which is realized in transition from partial derivatives to covariant ones.

*The fundamental consequence of the third criteria is the presence of the metric tensor $g_{\alpha\beta}$ in theory, which in case of absence of gravity turns into Minkowsky tensor $g_{\alpha\beta} \to \eta_{\alpha\beta}$.*

Further, one of the basic postulates of gravity theory is *the weak equivalence principle* (WEP):
*Trajectory of non-charged test body depends only on its initial position and initial velocity and does not depend on its intrinsic structure or composition.*

At present, WEP is verified in the ground-based conditions with the accuracy of $4 \cdot 10^{-13}$ (Will, 2001).

A kernel of metric theories reflects *Einstein equivalence principle (EEP)*. EEP includes WEP and states:
*The outcome of any local non-gravitational experiment does not depend on both the velocity of free falling laboratory, and the fact where and when in the Universe this experiment is performed.*

The theories consistent with EEP are referred to as metric ones. The structure of such theories is close to GR, but they can differ by the form of the field equations and by presence of additional fields, which do not interact directly with the matter.

Metric theories may be defined by the following postulates (Will, 1981, 2001):

*P1. Space-time is endowed with a metric $g_{\alpha\beta}$.*

*P2. World lines of test bodies are the geodesics of that metric.*

*P3. In local free falling reference frames, called as Lorentz frames, the non-gravitational laws of physics are those of special relativity.*

We note that the first two postulates have completely defined geometrical sense, i.e. they are pertinent to the mathematical model of space-time. The third postulate is formulated (as EEP) in physical terminology and requires on the mathematical representation. It is obvious that a mathematical model can not be deduced from the physical principle, and that a physical principle can not be equivalent to the system of mathematical postulates. In reality, the Lorentz reference frame is mathematically represented by tetrad, which is parallel transported along the geodesic world line of an observer. At that, the third postulate can be formulated as the following:

*P3'. Switching on gravitation field is mathematically represented by transition from pseudo-Euclidian connection of Minkowskian space-time to pseudo-Riemannian one, which corresponds to the metric $g_{\alpha\beta}$.*

In consequence of metricity condition $g_{\alpha\beta;\gamma} = 0$ the lengths of vectors and angles between them do not change by parallel transport.

## 3. Non-metrical generalizations of basics of relativistic gravity theory

Now we proceed to discuss generalizations in question. First of all, we note that our discussion concerns with only those gravity theories that are *geometric* ones. As in case of metric theories we accept that *4-dimensional manifold M is the model of space-time and gravitational interaction is described by a geometric structure of this manifold*. In this manner we preserve the universality of gravitational interaction.

Then, we require the fulfillment of *correspondence principle with GR*; to be exact we require that GR may be obtained as a result of continuous limiting procedure from a generalized theory, when some parameter $\varpi$ tends to zero. Main geometrical structures that are necessary for mathematical description of physical phenomena are the metric form and the connection. The first one enables us to create models of space-time measurements; the second one is another name for derivatives of tensor quantities. In Minkowsky space-time the metric is pseudo-Euclidian, and the connection refers to the absolute parallelism. In metric theories the connection is completely defined by Riemannian metric.

In the context of the mathematical model, we treat a test body as the mass point with which the commoving standard clock can be associated. According to WEP the test body motion is defined by the initial position and velocity. Note that there is a widespread conception that any complete gravitational theory, which agrees with WEP, must agree also with EEP (Schiff's conjecture) and therefore be metric theory. We suppose that future study of more advanced body models in the concerned framework has to confirm or disprove this hypothesis. However, at the moment we confine ourselves with examination of the point mass model.

Let $x^\mu = x^\mu(p)$ be a 4-trajectory of the body, $p$ is an arbitrary monotonous parameter. For proper time interval we put

$$d\tau^2 = G(x^\alpha, dx^\beta/dp)dp^2, \tag{1}$$

the function $G$ satisfying the homogeneity condition $G(x^\alpha, ky^\beta) = k^2 G(x^\alpha, y^\beta)$. The latter condition is necessary to provide parametric invariance of the proper time.

The function $G(x^\alpha, dx^\beta)$ defines Finsler geometry (Rund, 1959). But the only form (1) is not sufficient for modeling the space-time measurements. In order to do this one must have some analog of the metric tensor. From the homogeneity condition it follows that

$$d\tau^2 = G_{\mu\nu}(x^\alpha, dx^\beta)dx^\mu dx^\nu, \tag{2}$$

where

$$G_{\mu\nu}(x^\alpha, y^\beta) = \frac{1}{2}\frac{\partial^2 G(x^\alpha, y^\beta)}{\partial y^\mu \partial y^\nu}. \tag{3}$$

However, as distinct from the Riemannian case, for a given $d\tau$ tensor (3) is not unique to satisfy Eq. (2). Then we have not a good cause to admit the relation (3) as a definition of the metric tensor. If we introduce directly a symmetric tensor $G_{\mu\nu}(x^\alpha, y^\beta)$ of zero degree homogeneity in $y^\beta$ with rejecting (3), then we pass to the generalized Finsler geometries.

Thus we suppose that in our class of theories the following postulate takes place:

*1. On the space-time manifold $M$ there is a generalized Finsler metric tensor $G_{\mu\nu}(x^\alpha, y^\beta)$; and the interval squared between close events is presented by eq. (2); if $\varpi$ tends to zero, then $G_{\mu\nu}$ tends to the metric tensor of GR, $G_{\mu\nu}(x^\alpha, y^\beta) \xrightarrow{\varpi \to 0} g_{\mu\nu}(x^\alpha)$.*

According to the correspondence principle the metric tensor $G_{\mu\nu}(x^\alpha, y^\beta)$ at $\varpi = 0$ has a Lorentz signature. Because the signature may be changed only on hypersurfaces defined by the equation $\det[G_{\mu\nu}(\varpi, x^\alpha, y^\beta)] = 0$, we find that for any point $(x^\alpha, y^\beta)$ of the tangent bundle $TM$ there is a finite interval $\varpi \in I$, where the signature is preserved.

This is closely related to the following supposition:

*2. Equation $G(x_0^\alpha, dx^\beta) = 0$ defines a two-sheeted light cone in the tangent space $T_{x_0}M$ of the point $x_0$; there is a corresponding light ray for any of 4-directions satisfying to this equation.*

Then according to the weak equivalence principle, we put:

*3. There is a unique test body trajectory for any initial point $x_0^\alpha$ of space-time and for any 4-direction, such that $G(x_0^\alpha, dx^\beta) > 0$. We can relate to every trajectory a standard clock; the time counts of this clock are equal to the interval along the trajectory.*

At last, generalizing the postulate *P3'*, we do not yet introduce exactly the connection. But we require that

*4. The effect of gravitation on the matter dynamics is modeled by introduction of the non-trivial connection on $M$; at $\varpi = 0$ the connection is pseudo-Riemannian one corresponding to $g_{\alpha\beta}$.*

Eq. (1) fixes certain canonical parameterization of trajectories. It can be rewritten as

$$G(x^\alpha, \dot{x}^\beta) = 1, \tag{4}$$

and this may be considered as integral of motion or as some constraint. Therefore, only three components of 4-velocity are independent.

It follows from the third postulate that test body trajectories (i.e. generalized geodesics) satisfy a system of differential equations of the second order; they are analogous to the affine geodesics equations, however with the connection $K^{\alpha}_{\beta\gamma}(x^{\mu}, \dot{x}^{\nu})$ that depends on 4-velocity. We presume that these connection coefficients are of zero order homogeneous in the velocities. This is important by different reasons. Namely, it makes possible 1) to choose independently the measurement units of coordinate and proper time; 2) to extend these equations to light-like trajectories, when the parameter $\tau$ takes on a different physical meaning; 3) to extend these equations to space–like trajectories, when the interval squared becomes negative.

The consequence of this homogeneity is that the solutions of the generalized geodesic equation depend on initial velocities $\dot{x}^{\nu}|_{\tau=0}$ and the parameter only through the products $(\dot{x}^{\nu}|_{\tau=0} \cdot \tau)$.

If we consider the geodesics as auto-parallel curves (to have a generalized law of inertia) with respect to the physical connection $H^{\alpha}_{\beta\gamma}(x^{\mu}, \dot{x}^{\nu})$ (its existence follows from the 4-th postulate), then we have that it can be represented as:

$$H^{\alpha}_{\beta\gamma}(x^{\mu}, \dot{x}^{\nu}) = K^{\alpha}_{\beta\gamma}(x^{\mu}, \dot{x}^{\nu}) + L^{\alpha}_{\beta\gamma}(x^{\mu}, \dot{x}^{\nu}), \tag{5}$$

the tensor $L^{\alpha}_{\beta\gamma}$ satisfying the condition

$$L^{\alpha}_{\beta\gamma} \dot{x}^{\beta} \dot{x}^{\gamma} = 0. \tag{6}$$

Differentiating (4), we get the relation that connects generalized metric and geodesic structures:

$$\left(G_{\alpha\beta,\gamma} - 2G_{\alpha\delta} K^{\delta}_{\beta\gamma} - G_{\alpha\beta,,\delta} K^{\delta}_{\gamma\mu} \dot{x}^{\mu}\right) \dot{x}^{\alpha} \dot{x}^{\beta} \dot{x}^{\gamma} = 0. \tag{7}$$

The comma denotes the partial derivatives with respect to coordinates, and double comma represents the partial derivatives with respect to the velocity components. The latter equation is homogeneous of the third order in the velocities.

Then, the space-time geometry is defined by the metric tensor $G_{\mu\nu}(x^{\alpha}, y^{\beta})$ and the connection (5), which satisfies Eqs. (6), (7). Let $g_{\mu\nu}(x^{\alpha})$ be a GR metric tensor that corresponds to the same matter distribution, and let $\Gamma^{\alpha}_{\beta\gamma}(x)$ be corresponding to it Riemannian connection. According to the correspondence principle we admit that the deformation tensors $\Theta_{\mu\nu}(x^{\alpha}, y^{\beta}) = G_{\mu\nu}(x^{\alpha}, y^{\beta}) - g_{\mu\nu}(x^{\alpha})$ and $\Pi^{\alpha}_{\beta\gamma}(x^{\mu}, y^{\nu}) = H^{\alpha}_{\beta\gamma}(x^{\mu}, y^{\nu}) - \Gamma^{\alpha}_{\beta\gamma}(x^{\mu})$ are the first order values in $\varpi$.

**4. Problem of comparison with observation data**

As we have pointed out, the class of theories satisfying the above requirements contains a large set of different geometries. There is a question how to compare predictions of alternative theories with each other and with the experiment. For example, to compare two different spaces in differential geometry one typically assumes that they are defined in a common coordinate system. Does such a common coordinate system exist for the geometries under discussion?

The positive answer to these questions relies upon the following considerations. Practically the reference frames (RF) are determined by the optical observations of celestial bodies and certain operations with clocks, gyroscopes etc. Every complete theory of gravitation must include mathematical models of free test body motion, light propagation and these operations. The practical RF and corresponding theoretical model form an interface between empirical and theoretical worlds. Then we may take, as a common mathematical basis for the comparison, those definitions of RF that are founded just on these models. In GR such RF are well known. These are the so-called RF of local observer (Misner, Thorn, Wheeler, 1973; Pyragas et al., 1995), such as Fermi RF and optical RF (Synge, 1960). They are based on considerations of geodesics (e.g., light-like ones) and parallel transport of tetrad, and they can be used for generalization and comparison of different gravitational theories.

The exact analytic integration of the geodesic equation can be performed only in very simple spaces. But this problem has been solved by means of power series in a common case (Pyragas et al., 1995; Rund, 1959). Also, a technique for integration of geodesics and for construction of local observer RF is developed in weak field approximation (Alexandrov, Fedorova, 1999; Marzlin, 1994; Nesterov, 1999). In many instances for the comparison of results of alternative theories it is sufficient to consider geodesic deviations due to the tensor of connection deformation $\Pi^{\alpha}_{\beta\gamma}(x^{\mu}, y^{\nu})$. Let us introduce the normal coordinates $z^{\tau}$ corresponding to the metric $g_{\alpha\beta}$, and let the values at these coordinates be marked by star, e.g., $\overset{*}{g}_{\rho\sigma}(z^{\tau})$. Riemannian geodesics passing through the origin are straight lines $z^{\sigma} = v^{\sigma} \cdot \tau$ ($v^{\sigma}$ being initial 4-velocity). The connection deformation yields a geodesic deformation. Then we seek for the corresponding solution of deformed geodesics in the form of $z^{\sigma} = v^{\sigma}\tau + \zeta^{\sigma}(v^{\rho}\tau)$, and we confine ourselves to the linear approximation in $\varpi$. In **Appendix A** we show that the geodesic deformations are represented by an expression

$$\zeta^{\sigma}(v^{\rho}\tau) = -\tau \int_{0}^{\tau} \frac{\overset{*}{g}^{\sigma\mu}(v^{\rho}s)}{s^2} \left[ \int_{0}^{s} \frac{\overset{*}{\Pi}_{\mu}(v^{\rho}t)}{t} dt \right] ds, \quad (8)$$

where $\overset{*}{\Pi}^{\sigma}(v^{\rho}\tau) = \tau^2 \overset{*}{\Pi}^{\sigma}_{\mu\nu}(v^{\rho}\tau, v^{\kappa}\tau)v^{\mu}v^{\nu}$.

In case of the background Minkowsky metric and linear approximation to the Schwartzshild metric as a perturbation, Eq. (8) yields the well-known relation for deflection angle and gravitational time delay. This provides us a starting point to study possible effects of non-metric geometry perturbations on observable quantities and to find upper limits on deviations from metric theory. More general situations require investigations of deformation not only of geodesic lines but also of geometrical and external fields, connection, curvature tensor etc.

This work is supported by the Science and Technology Center in Ukraine (Grant NN43-2003).

**Appendix A**

Here we prove formula (8) for geodesic deformations. In the coordinate system that is normal for background Riemannian space, the equation of geodesics of deformed space is

$$\ddot{z}^{\sigma} + \left( \overset{*}{\Gamma}^{\sigma}_{\mu\nu}(z^{\rho}) + \overset{*}{\Pi}^{\sigma}_{\mu\nu}(z^{\rho}, \dot{z}^{\rho}) \right) \dot{z}^{\mu} \dot{z}^{\nu} = 0. \quad (A1)$$

In this coordinates the geodesics of the background space, passing through the origin, look like straight lines $z^{\sigma} = v^{\sigma}\tau$ and Riemannian connection coefficients satisfy a functional equation (Schouten, 1954)

$$\overset{*}{\Gamma}^{\sigma}_{\mu\nu}(z^{\rho}) z^{\mu} z^{\nu} = 0. \quad (A2)$$

We look for solutions of Eq. (A1) in the form of $z^{\sigma} = v^{\sigma}\tau + \zeta^{\sigma}(v^{\rho}\tau)$. In the linear approximation in $\varpi$, arguments of the deformation tensor are not disturbed: $\overset{*}{\Pi}^{\sigma}_{\mu\nu}(z^{\rho}, \dot{z}^{\kappa}) \to \overset{*}{\Pi}^{\sigma}_{\mu\nu}(v^{\rho}\tau, v^{\kappa}) = \overset{*}{\Pi}^{\sigma}_{\mu\nu}(v^{\rho}\tau, v^{\kappa}\tau)$. In the latter equality, the homogeneity of $\overset{*}{\Pi}^{\sigma}_{\mu\nu}(z^{\rho}, \dot{z}^{\kappa})$ in the second argument is taken into account. Also we take into account that Eq. (A2) yields

$$\overset{*}{\Gamma}^{\sigma}_{\mu\nu,\kappa}(v^{\rho}\tau)v^{\mu}v^{\nu}\zeta^{\kappa}\tau + 2\overset{*}{\Gamma}^{\sigma}_{\mu\nu}(v^{\rho}\tau)v^{\mu}\zeta^{\nu} = 0. \quad (A3)$$

Thus, we get an equation for deformation of geodesics:

$$\tau \ddot{\zeta}^{\sigma} + 2\overset{*}{\Gamma}^{\sigma}_{\mu\nu}(v^{\rho}\tau)v^{\mu}(\tau \dot{\zeta}^{\nu} - \zeta^{\nu}) = -\tau \overset{*}{\Pi}^{\sigma}_{\mu\nu} v^{\mu} v^{\nu}. \quad (A4)$$

If vector $\overset{*}{\Pi}^{\sigma}(v^{\rho}\tau) = \tau^2 \overset{*}{\Pi}^{\sigma}_{\mu\nu}(v^{\rho}\tau, v^{\kappa}\tau)v^{\mu}v^{\nu}$ and operator $D = \tau d/d\tau$ are introduced, then this equation takes on the form:

$$D(D\zeta^\sigma - \zeta^\sigma) + 2\overset{*}{\Gamma}{}^\sigma{}_{\mu\nu}(v^\rho\tau)v^\mu\tau(D\zeta^\nu - \zeta^\nu) = -\overset{*}{\Pi}{}^\sigma. \tag{A5}$$

Since deformed and background geodesics have the same initial velocity, the required solution has to satisfy zero initial conditions $\zeta^\sigma(0) = 0$, $\dot\zeta^\sigma(0) = 0$. Let us show that the solution is

$$\zeta^\sigma(v^\rho\tau) = -\tau\int_0^\tau \frac{\overset{*}{g}{}^{\sigma\mu}(v^\rho s)}{s^2}\left[\int_0^s \frac{\overset{*}{\Pi}{}_\mu(v^\rho t)}{t}dt\right]ds. \tag{A6}$$

When $s$ tends to zero, integrand $\dfrac{\overset{*}{g}{}^{\sigma\mu}(v^\rho s)}{s^2}\int_0^s \dfrac{\overset{*}{\Pi}{}_\mu(v^\rho t)}{t}dt$ tends to $\dfrac{1}{2}\overset{*}{\Pi}{}^\sigma{}_{\mu\nu}(0, v^\kappa)v^\mu v^\nu$. So, we see that integral is converging and initial conditions are fulfilled. Then, we have

$$\eta^\sigma \equiv D\zeta^\sigma - \zeta^\sigma = -\overset{*}{g}{}^{\sigma\mu}(v^\rho\tau)\int_0^\tau \frac{\overset{*}{\Pi}{}_\mu(v^\rho t)}{t}dt, \tag{A7}$$

$$D\eta^\sigma = -\overset{*}{\Pi}{}^\sigma(v^\rho\tau) + \left(D\overset{*}{g}{}^{\sigma\kappa}\right)\overset{*}{g}{}_{\kappa\mu}\eta^\mu = -\overset{*}{\Pi}{}^\sigma - \left(\tau v^\nu \overset{*}{g}{}^{\sigma\kappa}\frac{\partial \overset{*}{g}{}_{\kappa\mu}(\tau v^\rho)}{\partial(\tau v^\nu)}\right)\eta^\mu = -\overset{*}{\Pi}{}^\sigma - 2\tau\overset{*}{\Gamma}{}^\sigma{}_{\mu\nu}v^\nu\eta^\mu. \tag{A8}$$

In the latter equality we take into account symmetry of the tensor $\overset{*}{g}{}_{\kappa\nu,\mu}(z^\rho)z^\nu$, which is a consequence of the condition $\overset{*}{g}{}_{\kappa\nu}(z^\rho)z^\nu = \overset{*}{g}{}_{\kappa\nu}(0)z^\nu$ that is equivalent to (A2).

We see that expression (A6) indeed satisfies Eq. (A5).

Formula (A6) is valid for deformations due to disturbance of the metric as well as for affine and Finsler deformations of the connection. It becomes considerably simpler when the background metric is pseudo-Euclidean i.e. in case of weak gravitational fields.